\documentclass[a4paper,12pt]{article}
\pdfoutput=1

\usepackage{undertilde}

\usepackage{amssymb}
\usepackage{amsmath}
\usepackage{epsfig}
\usepackage{graphicx}

\makeatletter

\@addtoreset{equation}{section}
\makeatother
\def\be{\begin{equation}}
\def\ee{\end{equation}}
\def\bea{\begin{eqnarray}}
\def\eea{\end{eqnarray}}

\def\({\left(}
\def\){\right)}
\def\<{\left<}
\def\>{\right>}

\def\tr{{\mbox{tr}}}
\def\be{\begin{equation}}
\def\ee{\end{equation}}
\def\bea{\begin{eqnarray*}}
\def\eea{\end{eqnarray*}}
\def\ben{\begin{eqnarray}}
\def\een{\end{eqnarray}}
\def\({\left(}
\def\){\right)}
\def\<{\left<}
\def\>{\right>}
\def\!{\right|}
\def\|{\left|}

\def\[{\left[}
\def\]{\right]}

\def\+{\bar}
\def\mb{\mathbb}
\def\tr{{\mbox{tr}}}

\def\D{{\cal{D}}}
\def\L{{\cal{L}}}
\def\t{\widetilde}

\def\L{{\cal{L}}}

\def\eps{{\cal{\varepsilon}}}

\def\E{{\cal{E}}}

\def\l{{{\ell}}}

\def\h{\widehat}

\begin{document}

\setlength{\unitlength}{1mm}

\pagestyle{empty}
\vskip-10pt
\vskip-10pt
\hfill 
\begin{center}
\vskip 3truecm
{\Large \bf
M5 branes on $\mb{R}^{1,1} \times$Taub-NUT}
\vskip 2truecm
{\large \bf
Andreas Gustavsson}
\vspace{1cm} 
\begin{center} 
Physics Department, University of Seoul, Seoul 02504 KOREA
\end{center}
\vskip 0.7truecm
\begin{center}
(\tt agbrev@gmail.com)
\end{center}
\end{center}
\vskip 2truecm
{\abstract We study M5 branes on $\mb{R}^{1,1} \times$Taub-NUT that we view as a singular fibration. Reducing the M5 branes along the fiber gives 5d SYM. Due to the singularity, this 5d theory has a gauge anomaly and it is not supersymmetric. To cure these problems we add a supersymmetric gauged chiral WZW theory on the 2d submanifold where the circle fiber vanishes. In addition we add a mass term for the five scalar fields located on this submanifold. With all this, we obtain a fully supersymmetric and gauge invariant theory.}

\vfill
\vskip4pt
\eject
\pagestyle{plain}

\section{Introduction} 
Bosonization is the equivalence between a 2d theory of a Dirac fermion and a theory of a scalar field whose non-Abelian generalization is a WZW theory \cite{Witten:1983ar}. The partition function factorizes, which can be better understood by coupling the chiral part to a background gauge field, resulting in a chirally gauged WZW theory \cite{Witten:1991mm}. An analogous situation happens in 6d for the selfdual three-fom of the M5 brane. If we compute the partition function of a nonchiral three-form then it factorizes. This factorization can be better understood if one couples the chiral part to a background $C$-field that is a three-form gauge potential in 11d supergravity \cite{Witten:1996hc}. A somewhat different kind of chirally gauged WZW theory appears as the boundary theory of M2 branes \cite{Chu:2009ms}.

If we put the M5 brane on $\mb{R}^{1,1} \times $Taub-NUT and view the Taub-NUT space as a circle fibration over $\mb{R}^3$, then the radius of that circle vanishes at the origin, where we have the submanifold $\mb{R}^{1,1}$. This has a brane interpretation of a D4 brane intersecting with a D6 brane. Near the intersection there are open strings stretching between the two branes. More generally, on the intersection brane, which is $\mb{R}^{1,1}$, we have chiral fermions in the bifundamental of $U(N)\times U(Q)$ where in the brane picture there are $Q$ coincident D6 intersecting with $N$ coincident D4 branes. These chiral fermions have a gauge anomaly that cancels the corresponding gauge anomaly of the 5d SYM that lives on the D4 brane \cite{Dijkgraaf:2007sw}. By bosonizing we get a chirally gauged WZW on $\mb{R}^{1,1}$ and again that theory has a gauge anomaly that cancels the corresponding gauge anomaly of the D4 brane \cite{Witten:2009at}. The gauge anomaly of the D4 brane comes from a gravi-photon term in eq (\ref{gravi}). In this paper we will extend this analyzis to the supersymmetric case. We find that the combined system of 5d SYM on the D4 branes and supersymmetric chirally gauged supersymmetric WZW theories plus additional mass terms for the five scalar fields on $\mb{R}^{1,1}$ results in a supersymmetric and gauge invariant theory.

\section{The 5d super Yang-Mills}
We put the Abelian M5 brane on a circle bundle with the metric 
\bea
ds^2 &=& G_{\mu\nu} dx^{\mu} dx^{\nu} + r^2 \(d\psi + \kappa_{\mu} dx^{\mu}\)^2
\eea
Here $x^{\mu}$ are coordinates on the five-dimensional base manifold, $\psi \sim \psi + 2\pi$ is the fiber coordinate parametrizing the circle fiber with circumference $2\pi r$. The radius $r$ can depend on the base-manifold coordinates. The gravi-photon is denoted $\kappa_{\mu}$ and its curvature is denoted $w_{\mu\nu} = \partial_{\mu} \kappa_{\nu} - \partial_{\nu} \kappa_{\mu}$. We perform dimensional reduction along the circle fiber. That results in an Abelian 5d SYM. This has a non-Abelian generalization whose action is given by \cite{Linander:2011jy}
\bea
S &=& \int d^5 x \sqrt{-G} \frac{1}{4\pi^2 r} \L
\eea
where the Lagrangian density is
\bea
\L &=& - \frac{1}{4} F_{\mu\nu}^2 + \frac{r}{4} \eps^{\mu\nu\lambda\kappa\tau} \omega(A)_{\mu\nu\lambda} w_{\kappa\tau}\cr
&& - \frac{1}{2} (D_{\mu} \phi^A)^2 - \frac{m^2}{2} (\phi^A)^2 \cr
&& + \frac{i}{2} \bar\psi \Gamma^{\mu} D_{\mu} \psi - \frac{i}{4r} \bar\psi \Gamma^{\mu} \psi \partial_{\mu} r + \frac{i r}{16} \bar\psi \Gamma^{\mu\nu} \Gamma^{\h\psi} \psi w_{\mu\nu}\cr
&& - \frac{1}{2} \bar\psi \Gamma^A \Gamma^{\h\psi} [\psi,\phi^A] + \frac{1}{4} [\phi^A,\phi^B]^2
\eea
where a trace over the Lie algebra generators is understood and not written out explicitly for notational simplicity. The field content is a gauge field $A_{\mu}$, five scalar fields $\phi^A$ and a fermionic field $\psi$. We use here an 11d notation, where the gamma matrices are 11d. Likewise the fermionic field is an 11d Majorana spinor that is reduced to 6d where it is Weyl projected, and subsequently dimensionally reduced to 5d. The details are summarized in the Appendix \ref{Gamma}. The mass squared is given by the following rather complicated expression
\bea
m^2 &=& \frac{R}{5} - \frac{r^2}{20} w_{\mu\nu}^2 + \frac{3}{5} \frac{\nabla^2 r}{r} - \(\frac{\nabla_{\mu} r}{r}\)^2
\eea
and 
\bea
\omega(A)_{\mu\nu\lambda} &=& A_{\mu} \partial_{\nu} A_{\lambda} - \frac{2i}{3} A_{\mu} A_{\nu} A_{\lambda}
\eea 
is the Chern-Simons three-form, where it is understood that the indices $\mu,\nu,\lambda$ shall be antisymmetrized. We notice that the graviphoton term
\ben
\int d^5 x \sqrt{-G} \frac{1}{4\pi^2 r}  \frac{r}{4} \eps^{\mu\nu\lambda\kappa\tau} \omega(A)_{\mu\nu\lambda} w_{\kappa\tau} &=& \frac{1}{16\pi^2} \int dx^{\mu} \wedge dx^{\nu} \wedge dx^{\lambda} \wedge dx^{\kappa} \wedge dx^{\tau} \omega(A)_{\mu\nu\lambda} w_{\kappa\tau}\cr
&& \label{gravi}
\een
makes no reference to the 5d metric. The supersymmetry variations are
\bea
\delta \phi^A &=& i \bar\eps\Gamma^A\psi\cr
\delta A_{\mu} &=& i \bar\eps\Gamma_{\mu}\Gamma^{\h\psi}\psi\cr
\delta \psi &=& \frac{1}{2} \Gamma^{\mu\nu}\Gamma^{\h\psi}\eps F_{\mu\nu}\cr
&& + \Gamma^{\mu} \Gamma^A \eps \(D_{\mu} \phi^A + \frac{\partial_{\mu} r}{r} \phi^A\)\cr
&& + \frac{r}{2} \Gamma^A \Gamma^{\mu\nu} \Gamma^{\h\psi} \eps w_{\mu\nu} \phi^A\cr
&& - \frac{i}{2} \Gamma^{AB} \Gamma^{\h\psi} \eps [\phi^A,\phi^B]
\eea
where the supersymmetry parameter satisfies the following Killing spinor equation
\ben
\nabla_{\mu} \eps &=& M_{\mu} \eps\cr
M_{\mu} &=& \frac{1}{2r} \Gamma_{\mu}\Gamma^{\nu}\partial_{\nu}r - \frac{r}{8}\Gamma_{\mu}\Gamma^{\rho\sigma}\Gamma^{\h\psi}w_{\rho\sigma} - \frac{r}{4} w_{\mu\nu} \Gamma^{\nu}\Gamma^{\h\psi}\label{5dKSE}
\een
We may introduce a Weyl covariant derivative
\bea
\D_{\mu} \phi &=& D_{\mu} \phi^A + \frac{\partial_{\mu} r}{r} \phi^A\cr
\D_{\mu} \psi &=& D_{\mu} \psi + \frac{3}{2} \psi \frac{\partial_{\mu} r}{r} - \frac{1}{2} \Gamma_{\mu}{}^{\nu} \psi \frac{\partial_{\nu} r}{r}
\eea
In terms of this derivative the Lagrangian becomes
\bea
\L &=& - \frac{1}{4} F_{\mu\nu}^2 + \frac{r}{4} \eps^{\mu\nu\lambda\kappa\tau} \omega(A)_{\mu\nu\lambda} w_{\kappa\tau}\cr
&& - \frac{1}{2} (\D_{\mu} \phi^A)^2 + \frac{i}{2} \bar\psi \Gamma^{\mu} \D_{\mu} \psi\cr
&& + \frac{i r}{16} \bar\psi \Gamma^{\mu\nu} \Gamma^{\h\psi} \psi w_{\mu\nu}\cr
&& - \frac{1}{2} \bar\psi \Gamma^A \Gamma^{\h\psi} [\psi,\phi^A] + \frac{1}{4} [\phi^A,\phi^B]^2
\eea
We now notice that the complicated mass term for the scalars has got completely absorbed into the Weyl covariant derivative. The Weyl transformations act  as
\bea
G_{\mu\nu} &\rightarrow & e^{2\Omega} G_{\mu\nu}\cr
r &\rightarrow & e^{\Omega} r\cr
\phi^A &\rightarrow & e^{- \Omega} \phi^A\cr
\psi &\rightarrow & e^{- \frac{3}{2} \Omega} \psi\cr
A_{\mu} &\rightarrow & A_{\mu}
\eea
Under these transformations the Weyl covariant derivative transforms Weyl covariantly as
\bea
\D_{\mu} \phi^A &\rightarrow & e^{-\Omega} \D_{\mu} \phi^A\cr
\D_{\mu} \psi &\rightarrow & e^{-\frac{3}{2} \Omega} \D_{\mu} \psi
\eea
Using this, it can be easily seen that the action is Weyl invariant. In addition to this, one may also show that if we transform the supersymmetry parameter as
\bea
\eps &\rightarrow & e^{\frac{1}{2} \Omega} \eps
\eea 
then the supersymmetry variations are also Weyl invariant.

\section{Conditions on the geometry}
The Killing spinor equation (\ref{5dKSE}) is derived from the 6d conformal Killing spinor equation of the M5 brane by imposing the condition $\partial_{\psi} \eps = 0$. Solutions to the 6d conformal Killing spinor equation have been summarized in \cite{Baum}, but the condition $\partial_{\psi} \eps = 0$ was not considered there. 

The most general condition on the geometry from the Killing spinor equation is obtained by analyzing $[\nabla_{\mu},\nabla_{\nu}] \eps$. But not only will this lead to fairly complicated computations, but also we will not need this strong condition for our purposes here. For the purpose of checking supersymmetry of the action, we will only need the weaker conditions that arises from 
\ben
\Gamma^{\mu\nu} \nabla_{\mu} \nabla_{\nu} \eps &=& - \frac{R}{4} \eps\label{Lich}
\een
This equation limis our search for possible geometries, but since (\ref{Lich}) gives a weaker condition on the geometry than the Killing spinor equation itself, we will still need to show the existence of solutions to the Killing spinor equation. From (\ref{Lich}) we obtain the following conditions
\ben
20 \(\frac{\nabla_{\mu} r}{r}\)^2 - 8 \frac{\nabla^2 r}{r} + \frac{r^2}{4} w_{\mu\nu}^2 &=& R\label{inte1}\\
\nabla^{\nu} \(\frac{1}{r} w_{\nu\mu}\) + \frac{1}{4} \E_{\mu}{}^{\nu\lambda\rho\sigma} w_{\nu\lambda} w_{\rho\sigma} &=& 0\label{inte2}
\een
Now this geometry is best understood not as a five-manifold, but as the six-manifold that is a circle bundle over the base five-manifold. We may express the first integrability condition in terms of the curvature scalar of the six-manifold if we note the relation
\bea
R_{6d} &=& R - \frac{r^2}{4} w_{\mu\nu}^2 - 2\frac{\nabla^2 r}{r}
\eea
that relates the curvature scalars $R_{6d}$ of the six-manifold with the curvature scalar $R$ of the base five-dimensional base-manifold. By using this relation, the integrability condition (\ref{inte1}) can be expressed as 
\ben
R_{6d} &=& 20 \(\frac{\nabla_{\mu} r}{r}\)^2 - 10 \frac{\nabla^2 r}{r}\label{resulting constraint}
\een

\subsection{Conformally flat spacetimes}
The constraint (\ref{resulting constraint}) is satisfied for a conformally flat metric, where in order to preserve supersymmetry under dimensional reduction, we need to restrict such a metric to be on the form 
\bea
ds^2 &=& r^2 \eta_{\mu\nu} dx^{\mu} dx^{\nu} + r^2 d\psi^2
\eea
where $r = r(x^{\mu})$ does not depend on the fiber direction parametrized by $\psi \sim \psi + 2\pi$. To show this, we use the a standard formula for how the curvature scalar transforms under a Weyl rescaling $G_{\mu\nu} \rightarrow r^2 G_{\mu\nu}$, \cite{dos}
\bea
R &\rightarrow & \frac{1}{r^2} R - 2 (D-1) \frac{1}{r^3} G^{\mu\nu} \nabla_{\mu} \nabla_{\nu} r - (D-1)(D-4) \frac{1}{r^4} G^{\mu\nu} \nabla_{\mu} r \nabla_{\nu} r
\eea
for a $D$-dimensional manifold. Here we take $D =6$ and $G_{\mu\nu} = \eta_{\mu\nu}$. This formula then gives
\bea
R &\rightarrow & - 10 \frac{1}{r^3} \eta^{\mu\nu} \partial_{\mu} \partial_{\nu} r - 10 \frac{1}{r^4} \eta^{\mu\nu} \partial_{\mu} r \partial_{\nu} r
\eea
In order to see that this corresponds to the constraint (\ref{resulting constraint}), we need to express the expressions on the right-hand side that are to be computed with respect to the metric $G_{\mu\nu} = r^2 \eta_{\mu\nu}$ on the 5d base manifold, in terms of the flat matric $\eta_{\mu\nu}$. We have
\bea
(\nabla_{\mu} r)^2 &=& \frac{1}{r^2} \eta^{\mu\nu} \partial_{\mu} r \partial_{\nu} r\cr
\nabla^2 r &=& \frac{1}{r^2} \eta^{\mu\nu} \partial_{\mu} r \partial_{\nu} r + \frac{3}{r^3} \eta^{\mu\nu} \partial_{\mu} r \partial_{\nu} r
\eea
Using these results we find the result 
\bea
R &\rightarrow & 20  \(\frac{\nabla_{\mu} r}{r}\)^2 - 10 \frac{\nabla^2 r}{r}
\eea
for the curvature scalar of the Weyl transformed metric, in agreement with the integrability constraint (\ref{resulting constraint}).

This, however, is not sufficient to prove the existence of a Killing spinor, since we only study a weaker version of the integrability condition. But it is not difficult to construct Killing spinor solutions to (\ref{5dKSE}) explicitly for conformally flat metrics. The Killing spinor equation is
\bea
\partial_{\mu} \eps &=& \frac{1}{2r} \eps \partial_{\mu} r
\eea
It has the general solution
\bea
\eps &=& \sqrt{r} \xi
\eea
for a constant spinor $\xi$. This solution can also be obtained by starting from the flat metric $ds^2 = \eta_{\mu\nu} dx^{\mu} dx^{\nu} + d\psi^2$ and the Killing spinor $\xi$ and then making a Weyl rescaling $\xi \rightarrow \sqrt{r} \xi$.

\subsection{$\mb{R}^{1,1} \times$ multi-Taub-NUT}
Another class of six-manifolds that satisfy the constraints (\ref{inte2}) and (\ref{resulting constraint}) are of the form $\mb{R}^{1,1} \times X$ where we take $X$ to be a hyper-Kahler manifold with the metric 
\ben
ds^2 &=& U d\vec{x} \cdot d\vec{x} + \frac{1}{U} \(d\psi + \vec{\kappa} \cdot d\vec{x}\)^2\label{hyper}
\een
Here $\vec{x}$ parametrizes $\mb{R}^3$ and $U$ is a function on $\mb{R}^3$. Since $X$ is Ricci flat we have that $R_{6d} = 0$ and (\ref{resulting constraint}) reduces to 
\ben
\frac{2}{r^2} G^{\mu\nu} \nabla_{\mu} r \nabla_{\nu} r &=& \frac{1}{r} G^{\mu\nu} \nabla_{\mu} \nabla_{\nu} r\label{ricciflat}
\een
where the radius is $r = \frac{1}{\sqrt{U}}$. The left-hand side of (\ref{ricciflat}) is
\bea
\frac{1}{2 U^3} \vec{\nabla} U \cdot \vec{\nabla} U
\eea
and the right-hand side is
\bea
\frac{1}{2 U^3} \vec{\nabla} U \cdot \vec{\nabla} U - \frac{1}{2U^2} {\nabla}^2 U
\eea
We see that (\ref{ricciflat}) is satisfied if $U$ is harmonic everywhere on $\mb{R}^3$
\bea
{\nabla}^2 U &=& 0
\eea
except for points where $\frac{1}{U^2}$ vanishes. Let us now look at the constraint (\ref{inte2}). With the metric (\ref{hyper}) this constraint becomes
\ben
\partial_i (U w_{ij}) &=& 0\label{Uw}
\een
in Cartesian coordinates on $\mb{R}^3$. In addition, it is necessary for $w_{ij}$ to be closed everywhere outside the singular points. This leads to the solution 
\bea
w_{ij} &=& \epsilon_{ijk} \partial_k U
\eea 
which is automatically closed
\bea
\epsilon_{ijl} \partial_l w_{ij} = \partial_k \partial_k U = 0
\eea
outside singular points, and moreover it satisfies (\ref{Uw}),
\bea
\partial_i (U w_{ij}) = \epsilon_{ijk} \partial_i U \partial_k U + U \epsilon_{ijk} \partial_ i \partial_k U = 0
\eea
A general harmonic function on $\mb{R}^3$ has the form  
\bea
U &=& \frac{1}{R^2} + \frac{1}{2} \sum_{i=1}^N \frac{1}{|\vec{x}-\vec{x}_i|}
\eea
which leads to the multi Taub-NUT metric \cite{Sen:1997js}. There are singularities at $\vec{x} = \vec{x}_i$ for $i = 1,...,N$. One may notice that such singularities are fine since close to a singularity we have $\frac{1}{U^2} \nabla^2 U \sim r^2 \nabla^2 \frac{1}{r} = 0$. Let us now take a closer look at a singularity, starting with $N=1$ for which we have the Taub-NUT metric 
\bea
ds^2 &=& U dx^i dx^i + \frac{1}{U} \(d\psi + \kappa_i dx^i\)^2
\eea
where
\bea
U &=& \frac{1}{R^2} + \frac{1}{2 |\vec{x}|}
\eea
We can view the Taub-NUT space as a circle bundle over a base-manifold that is a rescaled version of $\mb{R}^3$ with the metric $G_{ij} = U \delta_{ij}$. If we use polar coordinates on $\mb{R}^3$ then the Taub-NUT metric is 
\bea
ds^2 &=& U \(dr^2 + r^2 \(d\theta^2 + \sin^2\theta d\varphi^2\)\) + \frac{1}{U}\(d\psi + \frac{1}{2} \cos\theta d\varphi\)^2
\eea
and the radius of the circle fiber is $1/\sqrt{U}$ where
\bea
U &=& \frac{1}{R^2} + \frac{1}{2 r}
\eea
The radius of the circle fibration (not to be confused with the radius $r$ of $\mb{R}^3$) vanishes at $r = 0$ so the circle fibration is singular. But the manifold is nonetheless smooth at $r = 0$. To see that, we may look at the metric close to $r = 0$. When $r << R^2$ we may neglect the constant term in $U$ so the metric is 
\bea
ds^2 &=& \frac{1}{2r} \(dr^2 + r^2 \(d\theta^2 + \sin^2\theta d\varphi^2\)\) + 2r\(d\psi + \frac{1}{2} \cos\theta d\varphi\)^2
\eea
The interpretation of this metric becomes clearer if we define 
\bea
r &=& \frac{\rho^2}{2}
\eea
Then the metric becomes
\ben
ds^2 &=& d\rho^2 + \frac{\rho^2}{4} \(d\theta^2 + \sin^2 \theta d\varphi^2\) + \frac{\rho^2}{4} \(2 d\psi + \cos\theta d\varphi\)^2\label{flatmetric}
\een
This is now the metric of flat $\mb{R}^4$. For the details of the construction of this flat metric we refer to Appendix \ref{flat}. The important point is that $\psi$ is $2\pi$ periodic. 

If we look at the Taub-NUT metric in the other limit when $r>>R^2$ we may approximate $U \approx \frac{1}{R^2}$ and the metric is 
\bea
ds^2 &=& \frac{1}{R^2} dx^i dx^i + R^2 \(d\psi + \kappa_i dx^i\)^2
\eea
which describes a cylinder of radius $R$. The Taub-NUT space thus interpolates between flat $\mb{R}^4$ at the origin and a cylinder at infinity. 

The graviphoton one-form is 
\bea
\kappa &=& \frac{1}{2} \cos\theta d\varphi
\eea
Its curvature two-form $w = d\kappa$ is 
\ben
w &=& - \frac{1}{2} \sin\theta d\theta \wedge d\varphi\label{w1}
\een
that is integrated over $S^2$ to 
\bea
\oint_{S^2} w &=& - 2\pi
\eea
Another way of expressing this curvature two-form is as
\bea
w_{ij} &=& \epsilon_{ijk} \partial_k U
\eea
in flat $\mb{R}^3$ with metric $\delta_{ij}$ where $\epsilon_{123} = 1$ and totally antisymmetric. It is also useful to express this same relation in a covariant form by using the metric $G_{ij} = U \delta_{ij}$ of the base. Covariantly we then have 
\ben
w_{ij} &=& \frac{1}{G^{1/6}} \eps_{ijk} G^{k\l} \partial_{\l} U\label{w2}
\een
where we define the covariant tensor $\eps_{ijk} = \sqrt{G} \epsilon_{ijk}$. We may now confirm the equivalence of the two expressions (\ref{w1}) and (\ref{w2}) by choosing polar coordinates on the base. We then need to study the expression
\bea
w_{\theta\varphi} &=& \frac{1}{G^{1/6}} \eps_{\theta\varphi r} G^{rr} \partial_r U
\eea
We start by rewriting everything in terms of $U$ using 
\bea
G &=& U^3\cr
G^{rr} &=& \frac{1}{U}
\eea
Then 
\bea
w_{\theta\varphi} &=& \frac{1}{U\sqrt{U}} \eps_{\theta\varphi r} \partial_r U
\eea
Next we notice that 
\bea
\partial_r U &=& - \frac{1}{2 r^2}
\eea
We then need to address the quesion of finding an explicit expression for the antisymmetric tensor component
\bea
\eps_{\theta\varphi r} = \sqrt{G} \epsilon_{\theta\varphi r} = U^{3/2} \epsilon_{\theta\varphi r}
\eea
Here $\epsilon_{\theta\varphi r} = r^2 \sin\theta$ is the determinant of the Jacobian when we go from Cartesian to Polar coordinates. We now have all ingredients. Putting them together, we obtain 
\bea
w_{\theta\varphi} = - \frac{1}{2} \sin\theta
\eea
which is in perfect agreement with (\ref{w1}).

For multi-Taub-NUT we take 
\bea
U &=& \frac{1}{R^2} + \frac{1}{2} \sum_{I=1}^N \frac{1}{|\vec{x} - \vec{x}_I|}
\eea
and the gravi-photon is implicitly defined through 
\bea
w_{ij}  &=& \eps_{ijk} \partial_k U
\eea
This is a sum of terms,
\bea
w_{ij} &=& \sum_i w_{ij}^I\cr
w_{ij}^I &=& \eps_{ijk} \partial_k U^I\cr
U^I &=& \frac{1}{2} \frac{1}{|\vec{x} - \vec{x}_I|}
\eea
which means that the gravi-photon itself is a sum of terms,
\bea
\kappa_i &=& \sum_I \kappa_i^I 
\eea
When $\vec{x}\approx \vec{x}_I$ the geometry is locally that of flat $\mb{R}^4$ if no other points $\vec{x}_J$ coincide with $\vec{x}_I$. If $Q_I$ points coincide at $\vec{x}_I$, then we have, locally near that point, the metric
\bea
ds^2 &=& \frac{Q}{2r} \(dr^2 + r^2 \(d\theta^2 + \sin^2 \theta d\varphi^2\)\) + \frac{2r}{Q} \(d\psi + \frac{Q}{2} \cos\theta d\varphi\)^2
\eea
If we put $r = \frac{\rho^2}{2}$ we get
\bea
\frac{1}{Q} ds^2 &=& d\rho^2 + \frac{\rho^2}{4} \(d\theta^2 + \sin^2\theta d\varphi^2\) + \frac{\rho^2}{4 Q^2} \(2 d\psi + Q \cos\theta d\varphi\)^2
\eea
If we put $\psi = Q\t\psi$ we get
\bea
\frac{1}{Q} ds^2 &=& d\rho^2 + \frac{\rho^2}{4} \(d\theta^2 + \sin^2\theta d\varphi^2\) + \frac{\rho^2}{4} \(2 d\t\psi + \cos\theta d\varphi\)^2
\eea
where $\t\psi \sim \t\psi + \frac{2\psi}{Q}$. This is the metric of the orbifold $\mb{C}^2/\mb{Z}_Q$ with the identification $(z_1,z_2) \sim (e^{2\pi i/Q} z_1,e^{2\pi i/Q} z_2)$. 

We conclude that the multi-Taub-NUT space $TN_N$ is everywhere smooth for $N$ non-coinciding singular points. But when $Q$ singular points coincide we get an orbifold singularity of the type $\mb{C}^2/\mb{Z}_{Q}$.

Having found these geometries of multi-Taub-NUT, it remains to estanblish the existence of Killing spinor solutions. To this end, we will simply review an argument from \cite{Witten:2009at} that shows that such six-manifolds support 8 real covariantly constant spinors. The existence of a covariantly constant spinor on $X$ implies that the Ricci tensor must vanish. For hyper-Kahler manifolds the Ricci tensor vanishes and the generic $SO(4) = SU(2)_+\times SU(2)_-$ holonomy group is reduced to $SU(2)_+$. According to the holonomy principle, a covariantly constant spinor is a singlet under the holonomy group $SU(2)_+$. Let us represent gamma matrices in the $SO(4)$ tangent space group of $X$ as 
\bea
\gamma^i &=& \sigma^i \otimes \sigma^1\cr
\gamma^4 &=& 1 \otimes \sigma^2
\eea
Then the embedding of the $SU(2)_{\pm}$ generators into $SO(4)$ is done as 
\bea
\sigma^i P_{\pm} &=& - \frac{i}{2} \gamma^{i4} \mp \frac{i}{4} \eps^{ijk} \gamma^{jk}
\eea
where $P_{\pm} = \frac{1}{2} \(1 \otimes 1 \pm 1 \otimes \sigma^3\)$. From this, we conclude that spinors that are not rotated by the holonomy group $SU(2)_+$ satisfy $P_+ \psi = 0$ so they are anti-Weyl spinors. The six-manifold has a tangent space group $SO(1,5)$. A Weyl spinor under this tangent space group has $4$ complex components and the constraint that it shall be invariant under $SU(2)_+$ imposes another Weyl projection leading to $2$ complex components. For the M5 brane there is in addition an $SO(5)$ R-symmetry group and the spinor has $4$ internal $SO(5)$ R-symmetry spinor components that leads to in total $2 \times 4 = 8$ complex spinor components, but there is a Majorana condition that one can impose on a spinor in $SO(1,5) \times SO(5) \subset SO(1,10)$ resulting in $8$ real components. These correspond to $8$ real supercharges \cite{Witten:2009at}. 

\section{Gauge and supersymmetry anomalies}
Let us assume that the the circle fiber vanishes on a two-dimensional submanifold $\Sigma$ with Lorentzian signature. Because the circle fibration degenerates on $\Sigma$, it is somewhat difficult to analyze what happens there directly. One way to circumvent this difficulty is by considering a tubular neighborhood around $\Sigma$. We then consider a five-manifold with a boundary four-manifold of the form $\Sigma \times S^2_{r_0}$ where $S^2_{r_0}$ is a small sphere that is enclosing $\Sigma$ and then in the end we shall take the limit $r_0\rightarrow 0$. There can be a magnetic flux through $S^2_{r_0}$, both for the gauge field and for the gravi-photon field. Of course one important difference is that for the gravi-photon the flux is fixed by the geometry, whereas for the gauge field we shall sum over all possible fluxes. Let us begin by assuming that the gauge group is Abelian. Then by allowing $r_0$ to take any value greater than zero, we find that these magnetic fluxes are produced by delta functions localized on $\Sigma$. We have the following modifications of the Bianchi identities,
\ben
\partial_{\mu} F_{\nu\lambda} + \partial_{\lambda} F_{\mu\nu} + \partial_{\nu} F_{\lambda\mu} &=& 2\pi Q_F \delta_{\mu\nu\lambda}\label{BiF}\\
\partial_{\mu} w_{\nu\lambda} + \partial_{\lambda} w_{\mu\nu} + \partial_{\nu} w_{\lambda\mu} &=& - 2\pi \sum_{I=I}^n Q^I \delta^I_{\mu\nu\lambda}\label{Biw}
\een
Here $Q^I$ denotes the charge at submanifold $\Sigma_I$. For the multi-Taub-NUT, these charges sum up to $N$
\bea
\sum_{I=I}^n Q^I &=& N
\eea
where $N = 1,2,3,...$ is the integer charactarizing the multi-Taub-NUT. Further, $\delta^I_{\mu\nu\lambda}$ denotes the Poincare dual of $\Sigma_I$, defined as
\ben
\int d^5 x \sqrt{-G} \eps^{\mu\nu\lambda\rho\sigma} \frac{1}{6} \delta^I_{\mu\nu\lambda} \frac{1}{2} \omega_{\rho\sigma} &=& \int_{\Sigma_I} d^2 \sigma \sqrt{-\eta} \eps^{\alpha\beta} \frac{1}{2} \omega_{\alpha\beta}\label{Poincaredual}
\een
for an arbitrary test-two-form $\omega_{\mu\nu}$. Here $\eta_{\alpha\beta}$ denotes the induced metric on $\Sigma$ that we parametrize by coordinates $\sigma^{\alpha}$.  

Now let us look for a possible six-dimensional origin of the magnetic charge $Q_F$. This would be an integral over a three-cycle,
\bea
Q_F &=& \frac{1}{2\pi} \int_{S^2_{r_0} \times S^1} H
\eea
where $S^1$ denotes the circle fiber. But since the tubular neighborhood was inserted by us by hand and can be shrunk to zero size, it does not really exist in the six-manifold as a genuine cycle. So there is no three-cycle $S^2_0 \times S^1$ in the six-manifold and that means that $Q_F = 0$.

There are two options for writing the gravi-photon term. Either
\ben
\L_{grav} &=& \frac{r}{4} \eps^{\mu\nu\lambda\rho\sigma} \omega(A)_{\mu\nu\lambda} w_{\rho\sigma}\label{g1}
\een
or 
\ben
\L'_{grav} &=& - \frac{r}{8} \eps^{\mu\nu\lambda\rho\sigma} F_{\mu\nu} F_{\lambda\rho} \kappa_{\sigma}\label{g2}
\een
The two ways differ by a total derivative,
\bea
\L_{grav} &=& \L'_{grav} + r\nabla_{\rho} \(\frac{1}{2} \eps^{\mu\nu\lambda\rho\sigma} \omega(A)_{\mu\nu\lambda} \kappa_{\sigma}\)
\eea
inconsequential for the equation of motion, but $\L'_{grav}$ is gauge invariant while $\L_{grav}$ is not, so one might prefer to use $\L'_{grav}$. But $\L'_{grav}$ is not invariant under a reparametrization of the fiber coordinate. Under a reparametrization $\psi \rightarrow \psi' = \psi + f$ the graviphoton field transforms as $\kappa_{\mu} \rightarrow \kappa'_{\mu} - \partial_{\mu} f$. So this way of writing the gravi-photon term is not invariant under such a 'geometric' gauge transformations. 

The gauge field equation of motion is
\bea
\nabla_{\nu} \(\frac{1}{r} F^{\nu\mu}\) + \frac{1}{4} \eps^{\mu\nu\lambda\rho\sigma} F_{\nu\lambda} w_{\rho\sigma} &=& 0
\eea
This equation of motion holds irrespectively of whether we use the gravi-photon term (\ref{g1}) or (\ref{g2}) in the action. If we act by $\nabla_{\mu}$ on the left-hand side, then we get
\ben
\frac{2}{r} R_{\mu\nu} F^{\mu\nu} + \frac{1}{4} \eps^{\mu\nu\lambda\rho\sigma} F_{\nu\lambda} \nabla_{\mu} w_{\rho\sigma} &=& 0 \label{eomA}
\een
The first term is zero because the Ricci scalar is symmetric while $F_{\mu\nu}$ is antisymmetric. But if $Q$ is nonzero, then the second term is not zero. In this case, the equation of motion (\ref{eomA}) becomes inconsistent, and must be modified somehow. Following \cite{Ohlsson:2012yn}, we make the following ansatz for such a modified equation of motion,
\ben
\nabla_{\nu} \(\frac{1}{r} F^{\nu\mu}\) + \frac{1}{4} \eps^{\mu\nu\lambda\rho\sigma} F_{\nu\lambda} w_{\rho\sigma} &=& J^{\mu}\label{eomAA}
\een
Now if we act on both sides by $\nabla_{\mu}$ we get
\bea
2\pi \eps^{\mu\nu\lambda\rho\sigma} \frac{1}{2} F_{\nu\lambda} \frac{1}{6}\sum_{I=1}^n Q^I \delta^I_{\mu\rho\sigma} &=& \nabla_{\mu} J^{\mu} 
\eea
and if the left-hand side is nonzero, then this shows that 5d SYM can not be the full story. Something more is needed that can produce $J_{\mu}$. 

Let us then use the gravi-photon term (\ref{g1}), which is not gauge invariant but diffeomorphism invariant. If we assume that the gauge group is non-Abelian, then under a finite gauge transformation
\ben
A^g &=& g^{-1} A_{\mu} g + i g^{-1} \partial_{\mu} g\label{gaugetransf5d}
\een
the Chern-Simons three-form transforms as
\ben
\omega(A^g)_{\mu\nu\lambda} &=& \omega(A)_{\mu\nu\lambda} + \partial_{\nu}\(i \partial_{\mu} g g^{-1} A_{\lambda}\) + \frac{1}{3} g^{-1} \partial_{\mu} g g^{-1} \partial_{\nu} g g^{-1} \partial_{\lambda} g\label{tt}
\een
The last term gives rise to the following term in the action,
\bea
\frac{1}{16\pi^2} \oint_{S^2} dx^i \wedge dx^j w_{ij} \int_{\mb{R}_+ \times \mb{R}^{1,1}} dx^{\mu} \wedge dx^{\nu} \wedge dx^{\lambda} \frac{1}{3} \tr \(g^{-1} \partial_{\mu} g g^{-1} \partial_{\nu} g g^{-1} \partial_{\lambda} g\) 
\eea 
Using that the magnetic charge is $-2\pi Q$ gives 
\bea
- \frac{Q}{12\pi} \int_{\mb{R}_+ \times \mb{R}^{1,1}} dx^{\mu} \wedge dx^{\nu} \wedge dx^{\lambda} \tr \(g^{-1} \partial_{\mu} g g^{-1} \partial_{\nu} g g^{-1} \partial_{\lambda} g\) 
\eea 
The manifold over which this is to be integrated is $\mb{R}_+ \times \mb{R}^{1,1}$ where $\mb{R}_+$ is the radial direction outwards from the $S^2$ over which we integrated $w_{ij}$. So this three-manifold has a boundary $S^2$. The metric on this manifold does not enter since the term is topological. Let us Wick rotate $\mb{R}^{1,1}$ into $\mb{R}^2$ that we subsequently compactify into $S^2$. Then the manifold over which we integrate has turned into $\mb{R}^3$ where we have removed a small ball at the center. Expecting nothing particular happens to the field $g$ at the origin, we can let this ball shrink to zero size and integrate over the full $\mb{R}^3$ space. Assuming that $g$ falls off sufficiently fast at infinity, this amounts to integrating over $S^3$ as we may then identify all points at infinity and make a one-point compactification of $\mb{R}^3$ to $S^3$ by adding the point at infinity. When integrating over $S^3$ this term is quantized in integer multiples of $2\pi i$ where the integer is the winding number as we map $S^3$ into an $SU(2)$ subgroup of the gauge group by the field $g$. 

For the second term in (\ref{tt}) we applying (\ref{Poincaredual}) after making an integration by parts. The final result is that the gravi-photon term transforms as
\bea
S_{grav}(A^g) &=& S_{grav}(A) - \sum_{I=1}^n \frac{i Q^I}{4\pi} \int_{\Sigma_I} dx^{\mu} \wedge dx^{\nu} \tr\(\partial_{\mu} g g^{-1} A_{\nu}\) - 2\pi n_w
\eea
where $n_w$ is an integer winding number. This integer plays no role in the quantum theory where one considers the exponentiated action $e^{i S}$ in Lorentzian signature.

The action is also not supersymmetric. But to study this problem it is advantagous to first consider a more general setup of a five-manifold with a generic four-manifold boundary. Then under a supersymmetry variation of the action, we will pick up the following boundary terms,
\bea
\delta S &=& \frac{1}{4\pi^2} \int d^5 x \partial_{\mu} \(\sqrt{-G} b^{\mu}\)\cr
b^{\mu} &=& - \frac{1}{r} F^{\mu\nu} \delta A_{\nu} - \frac{1}{4} \eps^{\mu\nu\lambda\rho\sigma} A_{\nu} \delta A_{\lambda} w_{\rho\sigma}\cr
&& - \frac{1}{r} D^{\mu} \phi^A \delta \phi^A - \frac{i}{2r} \bar\psi \Gamma^{\mu} \delta \psi
\eea
Explicitly we get
\bea
b^{\mu} &=& - \frac{i}{4 r} \bar\psi \Gamma^{\rho\sigma} \Gamma^{\mu} \eps \Gamma^{\h\psi} \eps F_{\rho\sigma} + \frac{i}{2r} \bar\psi \Gamma^{\rho} \Gamma^{\mu} \Gamma^A \eps D_{\rho} \phi^A\cr
&& - \frac{1}{4} \eps^{\mu\nu\lambda\rho\sigma} A_{\nu} i \bar\eps \Gamma_{\lambda} \Gamma^{\h\psi} \psi w_{\rho\sigma}\cr
&& - \frac{i}{2 r} \bar\psi \Gamma^{\mu} \Gamma^{\rho} \Gamma^A \eps \frac{\partial_{\rho} r}{r} \phi^A - \frac{i}{4} \bar\psi \Gamma^{\mu} \Gamma^{\rho\sigma} \Gamma^A \Gamma^{\h\psi} \eps w_{\rho\sigma} \phi^A\cr
&& - \frac{1}{4r} \bar\psi \Gamma^{\mu} \Gamma^{AB} \Gamma^{\h\psi} \eps [\phi^A,\phi^B]
\eea
We would now have liked to proceed along the lines of reference \cite{Belyaev:2008xk} and find boundary degrees that we add so that the total action becomes supersymmetric without imposing boundary conditions. This strategy works nicely for low-dimensional super Yang-Mills. But for 5d SYM this strategy fails. So instead we will use a different approach. In the end we are not interested in the 4d boundary theory, but in a 2d submanifold theory. So we want to take the limit where the radius $r_0$ goes to zero. As we are still integrating over $S^2_{r_0}$, taking the limit $r_0$ to zero, amounts to averaging over the radial directions. For most terms, such an averaging will produce zero net result because the fields are not expected to vary very much close to the singular point. From a 6d viewpoint, this singular point is perfectly regular and we expect the fields to be smooth close to the singularity. We then expect that the only terms that will survive the integration over $S^2_{r_0}$ will be those that arise as magnetic charges when we integrate the two-form $w_{\mu\nu}$. So we may isolate the terms that involve this two-form and only consider those terms. 

Let us analyze this problem on $\mb{R}^{1,1} \times TN$. We use the relation (\ref{wtor}) which leads to 
\bea
b^{\mu} &=& \frac{i}{8} \bar\psi \Gamma^A \Gamma^{\mu} \Gamma^{\rho\sigma} \Gamma^{\h\psi} \eps w_{\rho\sigma} \phi^A \cr
&& - \frac{1}{4} \eps^{\mu\nu\lambda\rho\sigma} i \bar\eps \Gamma_{\lambda} \Gamma^{\h\psi} \psi A_{\nu} w_{\rho\sigma} \cr
&& + ...
\eea
where we extracted terms proportional to $w_{\mu\nu}$. Now we use the fact that $w_{\mu\nu}$ has components only in the $\mb{R}^3$ base of $TN$ where it is a magnetic monopole of strength $- 2\pi Q$. The boundary is taken to be $\mb{R}^{1,1} \times S^2$ and the normal direction is the radial direction. Thus we are interested in the radial component
\bea
b^r &=& \frac{i}{4} \bar\psi\Gamma^A\Gamma^{r\theta\varphi} \Gamma^{\h\psi} \eps w_{\theta\varphi} \phi^A\cr
&& + \frac{1}{2} \E^{\mu\nu} \E^{r\theta\varphi} i \bar\psi \Gamma^{\h\psi} \Gamma_{\mu} \eps A_{\nu} w_{\theta\varphi}\cr
&& + ...
\eea
Now we need to examine the Weyl projections. On $TN$ we have the Weyl projection (\ref{WeylTN})
\bea
\Gamma^{123} \Gamma^{\h\psi} \eps &=& \eps
\eea
By combining that we the 6d Weyl projection
\bea
\Gamma^0 \Gamma^{123} \Gamma^4 \Gamma^{\h\psi} \eps &=& - \eps
\eea
we get
\bea
\Gamma^{04} \eps &=& - \eps
\eea
We define $\eps^{01234\h\psi} = \eps^{01234} = 1$, and we put $\eps^{\mu\nu ijk} = \eps^{\mu\nu} \eps^{ijk}$ where we put $\eps^{04} = - 1$ and $\epsilon^{123} = 1$. Then we we have
\bea
\epsilon^{\nu\lambda} \Gamma_{\nu} \eps &=& \Gamma^{\lambda} \eps
\eea
Using this relation, we get
\bea
b^r &=& \frac{i}{4} \E^{r\theta\varphi} \bar\psi\Gamma^A \eps w_{\theta\varphi} \phi^A\cr
&& + \frac{1}{2} \E^{r\theta\varphi} i \bar\psi \Gamma^{\h\psi} \Gamma^{\nu} \eps A_{\nu} w_{\theta\varphi}\cr
&& + ...
\eea
This leads to a variation of the action given by
\bea
\delta S_{5d} &=& - \sum_I \frac{Q_I}{8\pi} \int_{\Sigma_I} d^2 x i \bar\psi \Gamma^A \eps \phi^A\cr
&& - \sum_I \frac{Q_I}{4\pi} \int_{\Sigma_I} d^2 x i \bar\psi \Gamma^{\h\psi} \Gamma^{\mu} \eps A_{\mu}
\eea
In a 5d reduced notation
\ben
\delta S_{5d} &=& \sum_I \frac{Q_I}{8\pi} \int_{\Sigma_I} d^2 x i \chi \tau^A \E \phi^A\cr
&& - \sum_I \frac{Q_I}{4\pi} \int_{\Sigma_I} d^2 x \chi \gamma^{\mu} \E A_{\mu} \label{deltaS5d}
\een
where $\E^{\alpha\dot\alpha} = \eps^{\alpha - \dot\alpha}$ and $\chi^{\alpha\dot\alpha} = \psi^{\alpha + \dot\alpha}$. The 5d conjugate spinor is defined as $\bar\chi_{\beta\dot\beta} = \chi^{\alpha\dot\alpha} C_{\alpha\beta} C_{\dot\alpha\dot\beta}$. We further reduce to 2d notation by decomposing $\alpha = (u,m)$ and put $\gamma^{\mu} = (\gamma^{\mu})^u{}_v \delta^m_n$. Then $\chi \gamma^{\mu} \E = \chi_{u m \dot\alpha} (\gamma^{\mu})^u{}_v \E^{v m \dot\alpha}$. In the 2d reduced notation, the sum over $m$ is trivial and gives three identical copies. For more details on our spinor notations, we refer to Appendix \ref{Gamma}.

\section{WZW theories on $\Sigma_I$}
A supersymmetric WZW theory has been constructed in \cite{Abdalla:1984ef} using a superfield formulation. Here we will use a component formulation instead. Let us start by analyzing the following supersymmetric WZW Lagrangian (where $\eps^{01} = 1$),
\bea
\L_{WZW}(g,A_{\mu}) &=& \frac{1}{8\pi} \tr \(g^{-1} \(\partial_{\mu} - i A_{\mu}\) g\)^2 - \frac{i}{4\pi} \eps^{\mu\nu} \tr \partial_{\mu} g g^{-1} A_{\nu}\cr
&& - \frac{i}{8\pi} \tr\(\bar\lambda\gamma^{\mu}\partial_{\mu}\lambda\)\cr
&& + \frac{1}{12\pi}  \eps^{\mu\nu\lambda} \tr\(g^{-1}\partial_{\mu}g g^{-1}\partial_{\nu}g g^{-1}\partial_{\lambda}g\)
\eea
To get the action, we should integrated this over $\Sigma_I$, except for the last term that should be integrated over some three-manifold whose boundary is $\Sigma_I$. This Lagrangian is invariant under the supersymmetry variations 
\bea
g^{-1} \delta g &=& - \bar\eps\lambda\cr
\delta \lambda &=& - i \gamma^{\mu} \eps g^{-1} \partial_{\mu} g
\eea
provided we impose the following chiral projection on the supersymmetry parameter,
\bea
\gamma^{01} \eps &=& \eps
\eea
Under a variation of $A_{\mu}$, keeping $g$ and $\lambda$ fixed, we have 
\bea
\delta \L_{WZW}(g,A_{\mu}) &=& - \frac{i}{4\pi} \partial_{\mu} g g^{-1} \(\delta A^{\mu} + \eps^{\mu\nu} \delta A_{\nu}\) - \frac{1}{4\pi} A_{\mu} \delta A^{\mu}
\eea
If we make the specific variation 
\bea
\delta A_{\mu} &=& - \chi \gamma^{\mu} \E
\eea
as induced from 5d, then we get
\bea
\delta \L_{WZW}(g,A_{\mu}) &=& \frac{i}{4\pi} \partial_{\mu} g g^{-1} \chi \(\gamma^{\mu} + \eps^{\mu\nu} \gamma_{\nu}\) \E + \frac{1}{4\pi} A_{\mu} \chi \gamma^{\mu} \E
\eea
The first term vanishes by the projection $\gamma^{01} \E = \E$. So if we add the following 2d action,
\ben
S_{WZW} &=& \sum_I Q_I \int_{\Sigma_I} d^2 x \L_{WZW}(g_I,A_{\mu})\label{WZWaction}
\een
then we cancel the second term in (\ref{deltaS5d}).

The resulting Lagrangian is also not fully gauge invariant, where the gauge variation acts on the fields as
\ben
A_{\mu} &\rightarrow & h A_{\mu} h^{-1} - i \partial_{\mu} h h^{-1}\cr
g &\rightarrow & h g\label{gaugetransf2d}
\een
for a gauge parameter $h$. One obviously gauge noninvariant term is $\L_{non} = - \frac{i}{4\pi} \eps^{\mu\nu} \tr \partial_{\mu}g g^{-1} A_{\nu}$, which transforms into
\bea
\L_{non} &\rightarrow & \L_{non} - \frac{i}{4\pi} \eps^{\mu\nu} \tr\(h^{-1}\partial_{\mu}h A_{\nu}\)\cr
&& - \frac{1}{4\pi} \eps^{\mu\nu} \(\partial_{\mu} h h^{-1} \partial_{\nu} h h^{-1} + \partial_{\mu} g g^{-1} h^{-1} \partial_{\nu} h\)
\eea
The other gauge noninvariant term is the WZ term $\L_{WZ} = \frac{1}{12\pi}  \eps^{\mu\nu\lambda} \tr\(g^{-1}\partial_{\mu}g g^{-1}\partial_{\nu}g g^{-1}\partial_{\lambda}g\)$ that transforms into
\bea
\L_{WZ} &\rightarrow & \L_{WZ} + \frac{1}{12\pi} h^{-1}\partial_{\mu}h h^{-1} \partial_{\nu} h h^{-1} \partial_{\lambda} h\cr
&& + \frac{1}{4\pi} \eps^{\lambda\mu\nu} \partial_{\lambda} \(\partial_{\mu} h h^{-1} \partial_{\nu} h h^{-1} + \partial_{\mu} g g^{-1} h^{-1} \partial_{\nu} h\)
\eea
We now see that many terms cancel between $\delta \L_{non}$ and $\delta_{WZ}$ and we are left with the gauge variation
\bea
\delta S_{2d} &=& - \sum_I \frac{i Q_I}{4\pi} \int dx^{\mu} \wedge dx^{\nu} \tr\(h^{-1}\partial_{\mu}h A_{\nu}\)
\eea
In order to match the gauge variation (\ref{gaugetransf2d}) with the gauge transformation (\ref{gaugetransf5d}) we shall substitute $h$ here by $g^{-1}$, in which case we get
\bea
\delta S_{2d} &=& \sum_I \frac{Q_I}{4\pi} \int dx^{\mu} \wedge dx^{\nu} \tr\(i \partial_{\mu}g g^{-1}A_{\nu}\)
\eea
that is cancelling the gauge variation of $S_{5d}$.

\section{The WZW current}
If we would vary the gauge potential in the WZW theory, then we would get the equation of motion
\ben
\partial_{\mu} g g^{-1} - \eps_{\mu\nu} \partial^{\nu} g g^{-1} - i A_{\mu} &=& 0\label{EqA}
\een
This equation is not gauge invariant. But that is not so surprising. The gauge potential in the WZW theory is a background field and we are not supposed vary a background field. However, it is a dynamical field in the 5d SYM theory and so if we vary the gauge potential in the combined system of 5d SYM plus the WZW theory, then we should recover a gauge invariant equation of motion. The gauge invariant completion of the left-hand side in (\ref{EqA}) is  
\ben
D_{\mu} g g^{-1} - \eps_{\mu\nu} D^{\nu} g g^{-1}\label{EqB}
\een
Thus we are looking for a missing term $i \eps_{\mu\nu} A^{\nu}$ that should come from the 5d SYM upon variation of the gauge potential, so that (\ref{EqA}) is completed into (\ref{EqB}). Let us examine the gravi-photon term, and the following term
\bea
\frac{1}{16 \pi^2} A_{\mu} \partial_{\nu} \delta A_{\lambda} w_{\rho\sigma}
\eea
when we vary the gauge potential. To derive the Euler-Lagrange equation of motion, we would make an integration by parts. But here, when we make an integration by parts we must be careful because of (\ref{Biw}). It is not too hard to see that this will exactly produce our missing 2d term that will complete (\ref{EqA}) into (\ref{EqB}).

Let us follow \cite{Linander:2011jy} and put
\bea
J_{\mu} &=& D_{\mu} g g^{-1} - \eps_{\mu\nu} D^{\nu} g g^{-1}
\eea
Then the 5d SYM equation of motion is modified to 
\ben
D^{\nu} \(\frac{1}{r} F_{\nu\mu}\) + \frac{1}{4} \eps^{\mu\nu\lambda\rho\sigma} F_{\nu\lambda} w_{\rho\sigma} &=& \sum_I i \pi Q^I \delta^I_{123} J_{\mu}\label{EOMJ}
\een

Let us now assume the gauge group is Abelian for simplicity. Nothing essential changes in the argument we will make below when the gauge group is non-Abelian. When the gauge group is Abelian, we put 
\bea
g &=& e^{i \phi}
\eea
and then $\phi$ will be a $2\pi$ periodic scalar field, and the WZW action is given by 
\bea
S_{WZW} &=& - \frac{Q}{8\pi} \int d^2 x \((\partial_{\mu} \phi - A_{\mu})^2 + \eps^{\mu\nu} \phi F_{\mu\nu} + i \bar\lambda \gamma^{\mu} \partial_{\mu} \lambda\)
\eea
that is invariant under the supersymmetry variations
\bea
\delta \phi &=& i \bar\eps \lambda\cr
\delta \lambda &=& \gamma^{\mu} \eps (\partial_{\mu} \phi - A_{\mu})
\eea
for $\gamma \eps = \eps$. Under the gauge variation
\bea
\delta \phi &=& \Lambda\cr
\delta A_{\mu} &=& \partial_{\mu} \Lambda
\eea
we have the variation
\bea
\delta S_{WZW} &=& \frac{Q}{8\pi} \int d^2 x \Lambda \eps^{\mu\nu} F_{\mu\nu}
\eea
that is canceling against the gauge variation of the 5d SYM action,
\bea
\delta S_{5d} &=& - \frac{Q}{8\pi} \int d^2 x \Lambda \eps^{\mu\nu} F_{\mu\nu}
\eea
The equation of motion for $\phi$ is given by 
\ben
\nabla^2 \phi - \nabla^{\mu} A_{\mu} &=& \frac{1}{2} \eps^{\mu\nu} F_{\mu\nu}\label{scalarfieldEOM}
\een
For the gauge field, we should consider the combined system of 5d SYM coupled to 2d WZW, and then we find the equation of motion 
\bea
\nabla_{\nu} \(\frac{1}{r} F^{\nu\mu}\) + \frac{1}{4} \eps^{\mu\nu\lambda\rho\sigma} F_{\nu\lambda} w_{\rho\sigma} &=& - \pi Q \delta_{123} \(\nabla^{\mu} \phi - \eps^{\mu\nu} \nabla_{\nu}\phi - A^{\mu} + \eps^{\mu\nu} A_{\nu}\)
\eea
Acting on both sides by $\nabla_{\mu}$ we get
\bea
\frac{1}{4} \eps^{\mu\nu\lambda\rho\sigma} F_{\nu\lambda} \partial_{\mu} w_{\rho\sigma} &=& - \pi Q \delta_{123} \(\nabla^2 \phi - \nabla_{\mu} A^{\mu} + \frac{1}{2} \eps^{\mu\nu} F_{\mu\nu}\)
\eea
Consistency with (\ref{scalarfieldEOM}) implies that
\bea
\partial_i w_{jk} + \partial_k w_{ij} + \partial_j w_{ki} &=& - 2 \pi Q \delta_{123} 
\eea
which is a consistency check that everything fits together nicely.

In addition to this WZW theory, if we also add a mass term for the five scalar fields that is localized to $\Sigma_I$,
\bea
S_{mass} &=& - \frac{Q_I}{16 \pi} \int d^2 x \phi^A \phi^A
\eea
then that has the supersymmetry variation
\bea
\delta S_{mass} &=& - \frac{Q_I}{8\pi} \int d^2 x i \chi \tau^A \E \phi^A
\eea
that cancels the first term in (\ref{deltaS5d}).

\section{Discussion}
Our proposal is that we need to add additional degrees of freedom on $\Sigma_I$ that are not present in 5d SYM. A different viewpoint appears to have been taken in \cite{Ohlsson:2012yn} and \cite{Lambert:2018mfb} where it appears to have been suggested that we may obtain the WZW degrees directly from the degrees of 5d SYM. This latter viewpoint would then be similar with the idea that the KK modes are not supposed to be added but are already present in 5d SYM as solitonic solutions (instanton particles). 

An on-shell solution for the gauge field was constructed in \cite{Ohlsson:2012yn} by making use of the harmonic two-form on Taub-NUT. The strategy was to first solving the 6d equations of motion 
\bea
dH &=& 0\cr
H &=& * H
\eea
for the selfdual tensor field of the Abelian M5 brane on $\mb{R}^{1,1} \times TN$. The solution is given by
\bea
H &=& h_{\mu} dx^{\mu} \wedge \Omega
\eea
Here $\Omega$ is the unique antiselfdual harmonic two-form on $TN$,
\bea
\Omega &=& \frac{\partial_k U}{U^2} \(- e^k \wedge e^4 + \frac{1}{2} \eps^{ijk} e^i \wedge e^j\)
\eea
where
\bea
e^i &=& \sqrt{U} dx^i\cr
e^4 &=& \frac{1}{\sqrt{U}} \(d\psi + \kappa^i dx^i\)
\eea
This solution was  subsequently generalized to multi-Taub-NUT and shown to be supersymmetric in \cite{Lambert:2018mfb}.

Let us here return to the Taub-NUT space and let us take the gauge group to be Abelian, and let us make the following ansatz for an on-shell solution 
\bea
F_{i \mu} &=& \frac{\partial_i U}{U^2} h_{\mu}
\eea
with all other components vanishing. To see whether this satisfies the equation of motion, we shall start by computing 
\bea
\nabla^i \(\frac{1}{r} F_{i\mu}\)  + \frac{1}{2} \eps_{\mu}{}^{i \nu jk} F_{i \nu} w_{jk} &=&  - \frac{1}{U^{3/2}} \frac{1}{U^2} \partial_i U \partial_i U \(h_{\mu} + \eps_{\mu\nu} h^{\nu}\) + \frac{1}{U} \frac{1}{U^{3/2}}\partial_i \partial_i U h_{\mu}
\eea
The first term vanishes if we demand that
\bea
h_{\mu} + \eps_{\mu\nu} h^{\nu} &=& 0
\eea
The second term is identically zero. First we note that $\frac{1}{U^{3/2}} \partial_i \partial_i U$ is a delta function with respect to the integartion measure $d^3 x U^{3/2}$. Second, $1/U$ evaluated at $|x|=0$ is zero, which is killing the whole thing. Thus we find that this solves the equation of motion (\ref{EOMJ}) with $J_{\mu} = 0$. From the equation of motion $\partial^{\mu} F_{\mu i} = 0$ we find $\partial^{\mu} h_{\mu} = 0$. By combining this with $h_{\mu} = - \eps_{\mu\nu} h^{\nu}$ we get $\partial_{\mu} h_{\nu} - \partial_{\nu} h_{\mu} = 0$ that is locally solved by $h_{\mu} = \partial_{\mu} \varphi$ for some scalar field $\varphi$. Our computation shows a different result from \cite{Ohlsson:2012yn} where it was found that $J_{\mu} \sim h_{\mu}$, from which one would conclude that the scalar field $\phi$ in the WZW theory would be the same as the component $\varphi$ that sits in the super Yang-Mills field $F_{\mu i}$ rather than a new degree of freedom. We notice that there is no contradiction having $J_{\mu} = 0$ here since by acting on the equation of motion by $\partial^{\mu}$ we get zero on the left-hand side because $\partial_{\mu} w_{ij} + \partial_i w_{j \mu} + \partial_j w_{\mu i} = 0$ so we not generate a delta function from the left-hand side by acting by $\partial_{\mu}$ there. On the right-hand side we get of course zero too, if $J_{\mu} = 0$. 

In the reference \cite{Lambert:2018mfb} it was argued that we should use the gauge invariant gravi-photon term (\ref{g2}) rather than the Chern-Simons-like gravi-photon term (\ref{g1}) that we have used in this paper. But using (\ref{g2}) has the problem that it is not reparametrization invariant. If the 5d SYM would have no gauge anomaly, then there would be no need to add a gauged WZW theory and one could think this could then be the end of the story. However, the problem with supersymmetry would remain exactly the same as we have presented here. This is true regadless of what form of the gravi-photon term we choose as our preference. This makes us believe that it would be rather difficult to come up with some alternative construction that is based on (\ref{g2}). It should probably be possible, but one would then need to deal with a reparametrization anomaly and find some way to cancel this anomaly. The reference \cite{Lambert:2018mfb} also makes a couple of interesting observations. First, the solution found in \cite{Ohlsson:2012yn} was shown to be supersymmetric. It preserves all supersymmetries. There are no fermionic zero modes and no broken supersymmetries. The intersection brane was shown to carry an electric charge and it has a tension, both of which are expressed in terms of the function $h_{\mu}$ (denoted $\nu_+$ in \cite{Lambert:2018mfb}). This has a non-Abelian generaltzaion and also a generalization to multi-Taub-NUT \cite{Lambert:2018mfb}. It would be interesting to show that the mass saturates some BPS bound determined by some central charge. Presumably that central charge would be proportional to the electric charge. 

In \cite{Lambert:2018mfb} it was objected that the WZW for multi-Taub-NUT could not be the usual WZW with one three-manifold with $N$ different boundaries whenever $Q>1$. But there is no need for the three-manifold to connect all the $N$ different intersection branes. Some three-manifold could extend from one intersection brane out to infinity. Then if $Q$ intersection branes would coincide, then we just take the WZW level to be equal to $Q$ and extend the three-manifold to infinity, and it would not affect the WZW theories on the other intersection branes if we extend the three-manifold to infinity in such a way that it does not cross some of the other submanifolds. So by allowing the three-manifold to extend to infinity rather than joining different intersection branes, we seem to be able to avoid the problem that was raised in \cite{Lambert:2018mfb}. We notice that there is plenty of room to draw three-manifolds that extend to infinity. These three-manifolds are lines in $\mb{R}^3$ and extended along $\mb{R}^{1,1}$. Singular points where the circle fiber vanishes in multi-Taub-NUT corresponds to points in $\mb{R}^3$. So we can always find a line from any singular point in $\mb{R}^3$ that extends to infinity. There is no need to connect the different singular points with lines. These lines may instead extend to infinity and then we have genuinely different WZW theories for each singular point. But there appears to be some ambiguity in how we may choose to draw these lines though. We expect that different choices will lead to equivalent physical descriptions.

\section*{Acknowledgements}
This work was supported in part by NRF Grant 2020R1A2B5B01001473 and NRF Grant 2020R1I1A1A01052462.

\appendix
\section{Differential geometry}
\subsection{A formula for the spin connection}
Given a vielbein $e^a{}_{\mu}$, the spin connection $(\omega_{\mu})^a{}_b$ is implicitly defined by two equations. It is covariantly constant, and the torsion is vanishing,
\bea
\partial_{\mu} e^a{}_{\nu} + (\omega_{\mu})^a{}_b e^b{}_{\nu} - e^a{}_{\rho} \Gamma^{\rho}_{\mu\nu} &=& 0\cr
\Gamma^{\rho}_{[\mu\nu]} &=& 0
\eea
Then we get
\bea
(de^a)_{\mu\nu} + (\omega_{\mu})^a{}_b e^b{}_{\nu} - (\omega_{\nu})^a{}_b e^b{}_{\mu} &=& 0
\eea
We contract this equation by $e^{c\nu}$,
\bea
(de^{[d})_{\mu\nu} e^{c]\nu} + (\omega_{\mu})^{[dc]} + (\omega_{\nu})^{a[d} e^{c]\nu} e^a{}_{\mu} &=& 0\cr
\frac{1}{2} (de^a)_{\mu\nu} e^{c\mu} e^{d\nu} e^a{}_{\rho} + (\omega_{\mu})^{a[d} e^{c]\mu}e^a{}_{\rho} &=& 0
\eea
Subtracting these equations leaves us with 
\bea
(\omega_{\mu})^{dc} &=& e^{\nu[d} (\partial_{\mu} e^{c]}_{\nu} - \partial_{\nu} e^{c]}_{\mu}) - \frac{1}{2} e^{d\kappa} e^{c\tau} (\partial_{\kappa} e^a_{\tau} - \partial_{\tau} e^a_{\kappa}) e^a_{\mu}
\eea
This is a formula to compute the spin connection directly from the vielbein. 

\subsection{Covariant derivatives on a circle bundle}
We denote 6d objects with hats, 5d base manifold objects without the hat. So the 6d metric is 
\bea
ds^2 &=& \h{G}_{MN} dx^M dx^N
\eea
We assume this metric has the circle-bundle form
\bea
ds^2 &=& G_{\mu\nu} dx^{\mu} dx^{\nu} + r^2 \(d\theta + \kappa_{\mu} dx^{\mu}\)^2
\eea
The manifest Killing vector is $\h{v} = \partial_{\theta}$. For its lower components we have $\h{v}_{\theta} = r^2$ and $\h{v}_{\mu} = r^2 \kappa_{\mu}$. The vielbein is
\bea
\h{e}^{\h\theta} &=& r \(d\theta + \kappa_{\mu} dx^{\mu}\)\cr
\h{e}^{\h\mu} &=& e^{\h\mu}{}_{\nu} dx^{\nu}
\eea
The inverse vielbein is 
\bea
\h{e}_{\h\theta} &=& \frac{1}{r} \partial_{\theta}\cr
\h{e}_{\h\mu} &=& e^{\nu}{}_{\h\mu} \(\partial_{\nu} - \kappa_{\nu} \partial_{\theta}\)
\eea
The inverse metric is 
\bea
\h{G}^{MN} \partial_M \phi \partial_N \phi &=& \frac{1}{r^2} \partial_{\theta} \phi \partial_{\theta} \phi + G^{\mu\nu} (\partial_{\mu}\phi - \kappa_{\mu} \partial_{\theta}\phi) (\partial_{\nu}\phi - \kappa_{\nu} \partial_{\theta}\phi) 
\eea
If we perform dimensional reduction along the fiber, then we put $\partial_{\theta} = 0$ and we get
\bea
\h{G}^{MN} \partial_M \phi \partial_N \phi &=& G^{\mu\nu} \partial_{\mu} \phi \partial_{\nu} \phi
\eea
This corresponds to the metric 
\bea
ds^2 &=& G_{\mu\nu} dx^{\mu} dx^{\nu}
\eea
on the base-manifold. The simple form of this dimensionally reduced metric is why this form of the fiber-bundle metric is a preferred choice when we perform dimensional reduction.

The covariant derivatives of a vector field $\h\nabla_M \h{v}_N = \partial_M \h{v}_N - \h{\Gamma}_{MN}^P \h{v}_P$ are
\bea
\h\nabla_{\theta} v_{\theta} &=& \partial_{\theta} v_{\theta} - r \nabla^{\rho} r v_{\rho} - r \kappa_{\rho} \nabla^{\rho} r v_{\theta}\cr
\h\nabla_{\theta} v_{\mu} &=& \partial_{\theta} v_{\mu} - \frac{1}{2} r^2 w_{\mu}{}^{\rho} v_{\rho} + r \kappa_{\mu} \nabla^{\rho} r v_{\rho} \cr
&& + \frac{1}{2} r^2 w_{\mu}{}^{\rho} \kappa_{\rho} v_{\theta} - r \kappa_{\mu} \kappa_{\rho} \nabla^{\rho} r v_{\theta} - \frac{1}{r} \nabla_{\mu} r v_{\theta}\cr
\h\nabla_{\mu} v_{\nu} &=& \nabla_{\mu} v_{\nu} + r^2 w^{\rho}{}_{(\mu} \kappa_{\nu)} v_{\rho} + r \nabla^{\rho} r \kappa_{\mu} \kappa_{\nu} v_{\rho}\cr
&& - \nabla_{(\mu} \kappa_{\nu)} v_{\theta} - r^2 \kappa_{\rho} w^{\rho}{}_{(\mu} \kappa_{\nu)} v_{\theta} - r \kappa_{\rho} \nabla^{\rho} r \kappa_{\mu} \kappa_{\nu} v_{\theta} - \frac{2}{r} \nabla_{(\mu} r \kappa_{\nu)} v_{\theta}
\eea
The covariant derivatives of a spinor field $\h\nabla_M \psi = \partial_M \psi + \frac{1}{4} \omega_M^{\h{M}\h{N}} \Gamma_{\h{M}\h{N}} \psi$ are 
\bea
\h\nabla_{\theta} \psi &=& \partial_{\theta} \psi - \frac{r^2}{8} \Gamma^{\mu\nu} \psi w_{\mu\nu} - \frac{1}{2} \Gamma^{\mu} \Gamma^{\h\theta} \psi \partial_{\mu} r\cr
\h\nabla_{\mu} \psi &=& \nabla_{\mu} \psi + \kappa_{\mu} \(\h\nabla_{\theta}\psi - \partial_{\theta}\psi\) + \frac{r}{4} \Gamma^{\nu} \Gamma^{\h\theta} \psi w_{\mu\nu}
\eea
Here we define $\Gamma_{\mu} = \Gamma_{\h\nu} e^{\h\nu}{}_{\mu}$ and $\h\Gamma_{M} = \Gamma_{\h{N}} \h{e}^{\h{N}}{}_{M}$. 

\subsection{Reducing the conformal Killing spinor equation}
Let us now analyze the conformal Killing spinor equations
\bea
\h\nabla_{\mu} \eps &=& \h\Gamma_{\mu} \eta\cr
\h\nabla_{\theta} \eps &=& \h\Gamma_{\theta} \eta
\eea
We expand the covariant derivatives and the gamma matrices in 5d quantities,
\bea
\nabla_{\mu} \eps + \kappa_{\mu} \(\h\nabla_{\theta}\eps - \partial_{\theta}\eps\) + \frac{r}{4} \Gamma^{\nu} \Gamma^{\h\theta} \eps w_{\mu\nu} &=& \Gamma_{\mu} \eta + r \kappa_{\mu} \Gamma_{\h\theta} \eta\cr
\partial_{\theta} \eps - \frac{r^2}{8} \Gamma^{\mu\nu} \eps w_{\mu\nu} - \frac{1}{2} \Gamma^{\mu} \Gamma^{\h\theta} \eps \partial_{\mu} r &=& r \Gamma_{\h\theta} \eta
\eea
We put $\partial_{\theta} \eps = 0$ and get
\bea
\nabla_{\mu} \eps + \frac{r}{4} \Gamma^{\nu} \Gamma^{\h\theta} \eps w_{\mu\nu} &=& \Gamma_{\mu} \eta\cr
- \frac{r^2}{8} \Gamma^{\mu\nu} \eps w_{\mu\nu} - \frac{1}{2} \Gamma^{\mu} \Gamma^{\h\theta} \eps \partial_{\mu} r &=& r \Gamma_{\h\theta} \eta
\eea
We get
\bea
\nabla_{\mu} \eps + \frac{r}{4} \Gamma^{\nu} \Gamma^{\h\theta} \eps w_{\mu\nu} &=& - \frac{r}{8} \Gamma_{\mu} \Gamma^{\rho\sigma} \Gamma^{\h\theta} \eps w_{\rho\sigma} + \frac{1}{2r} \Gamma_{\mu} \Gamma^{\rho} \eps \partial_{\rho} r
\eea
that we may also write as
\bea
\nabla_{\mu} \eps &=& M_{\mu} \eps
\eea
where we define
\bea
M_{\mu} &=& \frac{1}{2r} \Gamma_{\mu} \Gamma^{\rho} \partial_{\rho} r  - \frac{r}{8} \Gamma_{\mu} \Gamma^{\rho\sigma} \Gamma^{\h\theta} w_{\rho\sigma} - \frac{r}{4} \Gamma^{\nu} \Gamma^{\h\theta} w_{\mu\nu}
\eea

\subsection{Covariantly constant spinors on Taub-NUT} 
The conditions that both the derivatives $\h\nabla_{\theta} \eps = 0$ and $\partial_{\theta} \eps = 0$ are vanishing imply that 
\ben
\frac{r^2}{4} \Gamma^{\mu\nu} \eps w_{\mu\nu} + \Gamma^{\mu} \Gamma^{\h\theta} \eps \partial_{\mu} r &=& 0\label{wtor}
\een
Let us study this condition in the context of Taub-NUT. There we have
\bea
\frac{1}{2} \Gamma^{ij} \eps w_{ij} = \frac{1}{U^{1/2}} \Gamma^i \Gamma^{\h\theta} \eps \partial_i U
\eea
We may express $w_{ij} = \epsilon_{ijk} \partial_k U$ in a covariant way as
\bea
w_{ij} &=& \frac{1}{\sqrt{U}} G^{k\l} \eps_{ijk} \partial_{\l} U
\eea
where $\eps_{ijk} = \sqrt{G} \epsilon_{ijk}$ is the covariant form of the antisymmetric tensor where $\eps_{123} = 1$ and totally antisymmetric. Then 
\bea
\frac{1}{2} \Gamma^{ij} \eps G^{k\l} \eps_{ijk} \partial_{\l} U &=& \Gamma^{\l} \Gamma^{\h\theta} \eps \partial_{\l} U
\eea
Let us cancel out $\partial_{\l} U$ on both sides and use $\Gamma_k \Gamma^k = 3$ to get
\ben
\frac{1}{6} \Gamma^{ijk} \eps \eps_{ijk} &=& \Gamma^{\h\theta} \eps\label{WeylTN}
\een
This is a Weyl projection condition that reduces the amount of supersymmetry by half and assures that $\eps$ is invariant under the $SU(2)_+$ holonomy group. 

Let us notice that for 
\bea
U &=& \frac{1}{R^2} + \frac{1}{2|\vec{x}|}
\eea
we have
\bea
w = - \frac{x_k}{4|\vec{x}|^3}  \epsilon_{ijk} dx^i \wedge dx^j
\eea
which is a monopole with charge $\oint_{S^2} w = - 2\pi$.

\section{Flat metric on $\mb{R}^4$}\label{flat}
The flat metric on $\mb{R}^4 = \mb{C}^2$ is $ds^2 = |dz_1|^2 + |dz_2|^2$. We may parametrize the space by Euler anges as
\bea
z_1 &=& \rho \cos\frac{\theta}{2} e^{- \frac{i}{2}(2 \psi+\varphi)}\cr
z_2 &=& \rho \sin\frac{\theta}{2} e^{- \frac{i}{2}(2 \psi-\varphi)}
\eea
where $\psi \sim \psi + 2 \pi$. These coordinates can be obtained by acting on the spin-up state with the Euler rotation
\ben
g &=& e^{- \frac{i}{2} \phi \sigma_3} e^{-\frac{i}{2} \theta \sigma_2} e^{- \frac{i}{2} 2\psi \sigma_3}\label{g}
\een
In this form it is clear that $g^{-1} = g^{\dag}$ so the rotation operator is unitary. The Maurer-Cartan forms are
\bea
g^{-1} dg &=& - \frac{i}{2} \sigma_a \omega^a
\eea
where 
\bea
\omega^1 &=& \sin 2\psi d\theta - \sin\theta \cos 2\psi d\phi\cr
\omega^2 &=& \cos 2\psi d\theta + \sin\theta \sin 2\psi d\phi\cr
\omega^3 &=& 2 d\psi + \cos\theta d\phi
\eea

\section{Gauge group normalization}
We assume the gauge group Lie algebra is 
\bea
[T_a,T_b] &=& i f_{ab}{}^c T_c
\eea
with the metric 
\bea
\tr(T_a T_b) = h_{ab} = \frac{1}{2} \delta_{ab}
\eea
We write the gauge potential as $A_{\mu} = A^a_{\mu} T_a$. The field strength is defined as
\bea
F_{\mu\nu} &=& \partial_{\mu} A_{\nu} - \partial_{\nu} A_{\mu} - i [A_{\mu},A_{\nu}]
\eea
or in component form
\bea
F^a_{\mu\nu} &=& \partial_{\mu} A^a_{\nu} - \partial_{\mu} A^a_{\nu} + f_{bc}{}^a A^b_{\mu} A^c_{\nu}
\eea
We fix the normalization by taking $SU(2)$ gauge group with group element $g$ as in (\ref{g}) parametrized by the Euler angles. Then we get
\bea
\tr(g^{-1} dg) &=& \frac{3}{2} \omega^1 \omega^2 \omega^3\cr
&=& - 3 \sin \theta d\theta d\varphi d \psi
\eea
which gives
\bea
\int \tr\(g^{-1} dg\)^3 &=& - 24 \pi^2
\eea
and consequently we shall normalize the Wess-Zumino term as
\bea
\frac{1}{12 \pi} \int \tr\(g^{-1} dg\)^3 &=& \frac{1}{12\pi} \int d^3 x \sqrt{-G} \eps^{\mu\nu\lambda} \tr\(g^{-1} \partial_{\mu} g g^{-1} \partial_{\nu} g^{-1} \partial_{\lambda} g\)
\eea
and it will be quantized in units of $2\pi$ where the integer measures the winding number as we map $S^3$ into $SU(2)$.

\section{Gamma matrices}\label{Gamma}
\subsection{Two dimensions}
In 2d we have the Majorana representation
\bea
\gamma^0 &=& i \sigma^2\cr
\gamma^1 &=& \sigma^1
\eea
The chirality matrix is then $\gamma = \gamma^{01} = \sigma^3$. The charge conjugation matrix is $C = \gamma^0 = i \sigma^2 = \eps$. The Dirac conjugate spinor is $\bar\psi = \psi^{\dag} \gamma^0$. Writing out the spinor components, we have the spinor $\psi^u$ acted on by the gamma matrices $(\gamma^{\alpha})^u{}_v$, however, the charge conjugatation matrix is $\eps_{uv}$ with component $\eps_{+-} = 1$. The Dirac conjugate is $\bar\psi_v = (\psi^u)^* i (\sigma^2)^u{}_v$. We note that $(\psi^u)^*$ transforms under Lorentz rotations like $\psi_u = \psi^v \eps_{vu}$. For the components, we have $\bar\psi_+ = - (\psi^-)^*$ and $\bar\psi_- = (\psi^+)^*$. We may impose the Majorana condition $\bar\psi_u = \psi^v \eps_{vu}$. In the Majorana representation it amounts to a spinor with real components, $(\psi^+)^* = \psi^+$ and $(\psi^-)^* = \psi^-$. We note that $C \gamma^{\mu}$ are symmetric. 
  
\subsection{Five dimensions}
We define 5d gamma matrices as
\bea
\gamma^0 &=& (\gamma^0)^u{}_v \delta^m_n\cr
\gamma^4 &=& (\gamma^1)^u{}_v \delta^m_n\cr
\gamma^i &=& \gamma^u{}_v (\sigma^i)^m{}_n
\eea
The charge conjugation matrix is
\bea
C &=& (\gamma^1)^u{}_v \eps_{mn}
\eea
We have
\bea
\gamma^{04} &=& \gamma^u{}_v \delta^m_n\cr
\gamma^{01234} &=& - i \delta^u_v \delta^m_n
\eea
We note that both $C$ and $C \gamma^{\mu}$ are antisymmetric, consistent with the decomposition of a product of two spinors into a scalar, a vector and an antisymmetric tensor,
\bea
4 \otimes 4 &=& 1_{anti} \oplus 5_{anti} \oplus 10_{symm} 
\eea

\subsection{Eleven dimensions}
We define the gamma matrices as
\bea
\Gamma^{\mu} &=& \gamma^{\mu} \otimes \sigma^1 \otimes 1\cr
\Gamma^{\psi} &=& 1 \otimes \sigma^2 \otimes 1\cr
\Gamma^A &=& 1 \otimes \sigma^3 \otimes \tau^A
\eea
where $A =6,7,8,9,10$ and $\tau^{12345} = 1$. The 6d chirality matrix is 
\bea
\Gamma = \Gamma^0 \Gamma^{123} \Gamma^4 \Gamma^{\psi} = 1 \otimes \sigma^3 \otimes 1
\eea
The charge conjugatation matrix is 
\bea
C &=& C_5 \otimes \eps \otimes C'_5
\eea
where $\eps = i \sigma^2$. Thus $C$ is antisymmetric while $C\Gamma^M$ for $M = (\mu,\psi)$ and $C\Gamma^A$ are symmetric. The Majorana condition is 
\bea
\bar\psi &=& \psi^T C
\eea
where $\bar\psi = \psi^{\dag} \Gamma^0$.

\section{5d SYM in reduced notation}
Expressed in terms of 5d gamma matrices and 5d spinors, we have 
\bea
\delta \phi^A &=& - i \E \tau^A \chi\cr
&=& i \chi \tau^A \E\cr
\delta A_{\mu} &=& \E \gamma_{\mu} \chi\cr
&=& - \chi\gamma_{\mu} \E\cr
\delta \chi &=& - \frac{i}{2} \gamma^{\mu\nu} \E F_{\mu\nu} - \gamma^{\mu} \tau^A \E \D_{\mu} \phi^A\cr
&& - \frac{i r}{2} \gamma^{\mu\nu} \tau^A \E w_{\mu\nu} \phi^A - \frac{1}{2} \tau^{AB} \E [\phi^A,\phi^B]
\eea


\begin{thebibliography}{999}
\bibitem{Sen:1997js}
A.~Sen,
``Dynamics of multiple Kaluza-Klein monopoles in M and string theory,''
Adv. Theor. Math. Phys. \textbf{1}, 115-126 (1998)
[arXiv:hep-th/9707042 [hep-th]].

\bibitem{Baum}
H.~Baum, F.~Leitner, ``The twistor equation in Lorentzian spin geometry,'' Math. Z. 247 (2004) 795.



\bibitem{Linander:2011jy}
H.~Linander and F.~Ohlsson,
``(2,0) theory on circle fibrations,''
JHEP \textbf{01} (2012), 159
[arXiv:1111.6045 [hep-th]].


\bibitem{Ohlsson:2012yn}
F.~Ohlsson,
``(2,0) theory on Taub-NUT: A note on WZW models on singular fibrations,''
[arXiv:1205.0694 [hep-th]].


\bibitem{Witten:2009at}
E.~Witten,
``Geometric Langlands From Six Dimensions,''
[arXiv:0905.2720 [hep-th]].



\bibitem{dos}
dos Santos, Wytler Cordeiro. (2020). ``Notes on the Weyl tensor, decomposition of Riemann tensor, Ruse-Lanczos identity and duality of the curvature tensor,'' Zenodo. https://doi.org/10.5281/zenodo.3814369


\bibitem{Belyaev:2008xk}
D.~V.~Belyaev and P.~van Nieuwenhuizen,
``Rigid supersymmetry with boundaries,''
JHEP \textbf{04} (2008), 008
[arXiv:0801.2377 [hep-th]].



\bibitem{Witten:1991mm}
E.~Witten,
``On Holomorphic factorization of WZW and coset models,''
Commun. Math. Phys. \textbf{144} (1992), 189-212


\bibitem{Witten:1983ar}
E.~Witten,
``Nonabelian Bosonization in Two-Dimensions,''
Commun. Math. Phys. \textbf{92}, 455-472 (1984)

\bibitem{Dijkgraaf:2007sw}
R.~Dijkgraaf, L.~Hollands, P.~Sulkowski and C.~Vafa,
``Supersymmetric gauge theories, intersecting branes and free fermions,''
JHEP \textbf{02}, 106 (2008)
[arXiv:0709.4446 [hep-th]].

\bibitem{Witten:1996hc}
E.~Witten,
``Five-brane effective action in M theory,''
J. Geom. Phys. \textbf{22}, 103-133 (1997)
[arXiv:hep-th/9610234 [hep-th]].

\bibitem{Lambert:2018mfb}
N.~Lambert and M.~Owen,
``Charged Chiral Fermions from M5-Branes,''
JHEP \textbf{04} (2018), 051
[arXiv:1802.07766 [hep-th]].


\bibitem{Ohlsson:2012yn}
F.~Ohlsson,
``(2,0) theory on Taub-NUT: A note on WZW models on singular fibrations,''
[arXiv:1205.0694 [hep-th]].


\bibitem{Abdalla:1984ef}
E.~Abdalla and M.~C.~B.~Abdalla,
``Supersymmetric Extension of the Chiral Model and {Wess-Zumino} Term in Two-dimensions,''
Phys. Lett. B \textbf{152}, 59-62 (1985)

\bibitem{Chu:2009ms}
C.~S.~Chu and D.~J.~Smith,
``Multiple Self-Dual Strings on M5-Branes,''
JHEP \textbf{01}, 001 (2010)
[arXiv:0909.2333 [hep-th]].


\end{thebibliography}
\end{document}